\documentclass[twocolumn,aps,amsmath,amssymb]{revtex4-2}
\usepackage[T1]{fontenc}
\usepackage{graphicx}
\usepackage{braket}
\usepackage{hyperref} 
\hypersetup{colorlinks,citecolor=red,urlcolor=blue}

\begin{document}

\title{Cascaded variational quantum eigensolver algorithm}

\author{Daniel Gunlycke}
\email{lennart.d.gunlycke.civ@us.navy.mil}
\author{C. Stephen Hellberg}
\author{John P. T. Stenger}
\affiliation{U.S. Naval Research Laboratory, Washington, DC 20375, United States}

\begin{abstract}
We present a cascaded variational quantum eigensolver algorithm that only requires the execution of a set of quantum circuits once rather than at every iteration during the parameter optimization process, thereby increasing the computational throughput.  This algorithm uses a quantum processing unit to probe the needed probability mass functions and a classical processing unit perform the remaining calculations, including the energy minimization.  The ansatz form does not restrict the Fock space and provides full control over the trial state, including the implementation of symmetry and other physically motivated constraints.
\end{abstract}

\maketitle

Quantum computing (QC) offers inherent advantages over classical computing (CC) for solving certain mathematical tasks~\cite{Benioff80,Feynman82,Deutsch85,Shor94,Shor97,Grover96,Lloyd96,Harrow09}.  One of the most promising application areas is the simulation of quantum-mechanical systems~\cite{Lloyd96,Abrams97,Ortiz01}.  Because the dimension of the Hilbert space that comprises the quantum states of a fermionic system increases exponentially with the system size, performing operations on this space is an intractable task for conventional classical computers, for all but the smallest systems.  A quantum computer, on the other hand, can process such a Hilbert space by mapping it to the Hilbert space of a quantum register---the size of which increases exponentially with the number of qubits---and then performing quantum gate operations on this register.

The two main algorithms for QC calculations of quantum-mechanical systems are the quantum phase estimation algorithm~\cite{Kitaev96} and the variational quantum eigensolver (VQE) algorithm~\cite{Peruzzo14}.  By recruiting classical computers for computationally efficient tasks, the latter algorithm requires relatively few gate operations, which limits the decoherence during the computations.  As less exposure to decoherence allows for higher computational fidelities, this algorithm has a reduced need for quantum error correction, making it ideal for current noisy intermediate-scale quantum (NISQ) computing~\cite{Preskill18}.  Since its introduction, the VQE algorithm has been applied to calculate the ground-state energy of a number of systems in chemistry and physics~\cite{Peruzzo14,McClean16,OMalley16,McClean17,Li17,Kandala17,Colless18,Wang19,Arute19,Parrish19,Fischer19,McArdle20,Seki20,Huggins20,Arute20,Endo21,Cerezo21,Zhang22,Gonthier22,Tilly22,Zhao22}.

One downside of the VQE algorithm is that the computational throughput is limited by the large number of needed quantum circuit executions on the quantum processing unit (QPU).  For each energy minimization, this number is the product of the number of nonzero coefficients in the Pauli expansion of the Hamiltonian that describes the system times the number of shots in the sampling process times the number of iterations in the optimization process times the number of energy values needed per iteration in the chosen optimization routine.  The limitation is in part caused by the dependence of the quantum circuits on the variational parameters, which intertwines the sampling and optimization processes and requires that the quantum circuits be executed again every time the parameters are updated.

\begin{figure}[t]
	\centering
	\includegraphics[width=\columnwidth]{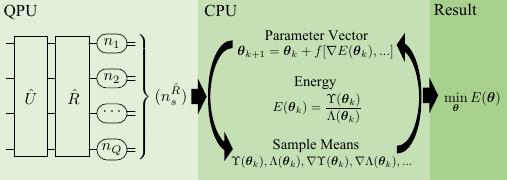}
	\caption{Schematic of an implementation of the cascaded variational quantum eigensolver algorithm.  The QPU executes a set of quantum circuits, each generating a unique quantum state $\hat R\hat U\!\ket{0}$ that when measured yield a family of occupation numbers $(n_1, n_2, ..., n_Q)$ recorded as $n_s$.  Repeating the same measurements multiple times for different $\hat R$ produces collections of families $(n_s^{\hat R})$ that are passed on as input to the CPU.  The CPU uses these samples together with a parameter vector $\boldsymbol{\theta}_k$ to compute derivatives of the energy $E(\boldsymbol{\theta})$ of the trial state $\ket{\Psi(\boldsymbol{\theta})}$ at $\boldsymbol{\theta}=\boldsymbol{\theta}_k$ by obtaining sample means for $\Upsilon(\boldsymbol{\theta}_k)$, $\Lambda(\boldsymbol{\theta}_k)$, $\nabla\Upsilon(\boldsymbol{\theta}_k)$, and $\nabla\Lambda(\boldsymbol{\theta}_k)$ in Eq. (\ref{e.14}) and (\ref{e.21}).  These derivatives are then used to generate a new parameter vector $\boldsymbol{\theta}_{k+1}$, using some optimization method $f[\nabla E(\boldsymbol{\theta}_k),...]$, and the process is repeated until the optimization has been completed and the sought minimum energy obtained.}
	\label{f.1}
\end{figure}
To address this challenge, we propose the cascaded variational quantum eigensolver (CVQE) algorithm, in which the variational parameters are exclusively processed on the classical processing unit (CPU).  The QPU is still needed to implement and measure a guiding state to yield probability mass functions that are then used in the optimization process.  This approach is possible because even though the dimension of the Hilbert space increases exponentially with the system size, the number of variational parameters in any VQE algorithm can at most increase polynomially---or else the amount of needed computing resources would grow exponentially.  Another benefit of the separation of the quantum circuit executions on the QPU and the optimization process on the CPU is that the optimization process in the CVQE algorithm partly compensates for the errors introduced during the quantum circuit executions.

As illustrated in Fig.~\ref{f.1}, given the samples from an initial set of measurements on the QPU, the energy minimization can subsequently be completed on the CPU alone.  By breaking the back-and-forth between the QPU and CPU in every iteration of the optimization process in the VQE algorithm, the CVQE algorithm reduces the number of quantum circuit executions by the factor of the number of energy values that needs to be calculated during the optimization process.  For instance, consider an optimization process that estimates the gradient using the simultaneous perturbation stochastic approximation~\cite{Spall92}, which requires 2 energy values per iteration, and needs 250 iterations (cf. the calculations of BeH$_2$ in Ref.~\cite{Kandala17}).  The computational throughput using the CVQE algorithm would in this case be increased by a factor of 500.  In other words, we could now complete calculations that would previously have taken months in a matter of hours.

In order to demonstrate the method behind the CVQE algorithm, consider a system of identical fermions described by the Hamiltonian $\hat H$ and let the antisymmetric Fock space $\mathcal{F}$ serve as the representation space for the quantum states of this system.  Our goal is to get an upper bound for the ground-state energy $E_\text g$ of the system by applying the variational method of quantum mechanics, which can be stated as
\begin{equation}
	E_\text g \le \min_{\boldsymbol{\theta}} E(\boldsymbol{\theta}),
	\label{e.1}
\end{equation}
where $\boldsymbol{\theta}$ is a variational parameter vector in the parameter space $\Theta$, which is a subset of the $d$-dimensional real coordinate space $\mathbb R^d$, and $E(\boldsymbol{\theta})$ is the energy of the trial state $\ket{\Psi(\boldsymbol{\theta})}$ in the ansatz.

We construct the trial state $\ket{\Psi(\boldsymbol{\theta})}$ from the normalized guiding state
\begin{equation}
	\ket{\Psi_0}=\hat U\ket{0},
	\label{e.2}
\end{equation}
where $\hat U$ is a unitary operator and $\ket{0}$ is the vacuum state in $\mathcal F$, that we prepare on the QPU for sampling.  In contrast to the unitary operators applied on the QPU in the commonly used unitary coupled cluster ansatz~\cite{Yung14,Peruzzo14,McClean16,OMalley16,Shen17,Harsha18,Kivlichan18,Hempel18,Romero18,Lee19,Dallaire19,Setia19,Grimsley20,Matsuzawa20,Sokolov20,Nam20,Motta21,Bauman21}, the hardware-efficient ansatz~\cite{Kandala17,Barkoutsos18,Kandala19,Ganzhorn19,Kokail19,Nakanishi19,Gard20,Bravo20,Tkachenko21}, and those used in various adaptive or trainable VQE algorithms~\cite{Grimsley19,Rattew19,Chivilikhin20,Bilkis21,Cincio21,Tang21,Yordanov21,ZhangS22,Du22}, we require that $\hat U$ be independent of $\boldsymbol{\theta}$.  Instead, we introduce the dependence on $\boldsymbol{\theta}$ through the operator $e^{i\hat\lambda(\boldsymbol{\theta})}$ that transforms $\ket{\Psi_0}$ to our trial state
\begin{equation}
 	\ket{\Psi(\boldsymbol{\theta})} = e^{i\hat\lambda(\boldsymbol{\theta})}\ket{\Psi_0},
	\label{e.3}
\end{equation}
where $\hat\lambda(\boldsymbol{\theta})$ is an operator.  Consequently, the energy of $\ket{\Psi(\boldsymbol{\theta})}$ is of the form
\begin{equation}
	E(\boldsymbol{\theta}) = \frac{\Upsilon(\boldsymbol{\theta})}{\Lambda(\boldsymbol{\theta})},
	\label{e.4}
\end{equation}
with the expectation values
\begin{subequations}
	\begin{align}
		\Upsilon(\boldsymbol{\theta}) =& \braket{\Psi_0|e^{-i\hat\lambda^\dagger(\boldsymbol{\theta})}\hat He^{i\hat\lambda(\boldsymbol{\theta})}|\Psi_0},\label{e.5a}\\
		\Lambda(\boldsymbol{\theta}) =& \braket{\Psi_0|e^{-i\hat\lambda^\dagger(\boldsymbol{\theta})}e^{i\hat\lambda(\boldsymbol{\theta})}|\Psi_0}.\label{e.5b}
	\end{align}
	\label{e.5}%
\end{subequations}

\begin{figure}[t]
	\centering
	\includegraphics[width=\columnwidth]{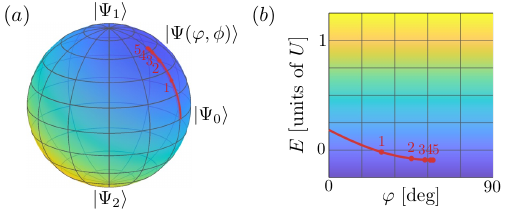}
	\caption{The energy $E(\varphi,\phi)$ of the singlet two-electron trial state $\ket{\Psi(\varphi,\phi)}$ of the two-site Hubbard model with the Hamiltonian $\hat H=t\sum_\sigma\big(c_{0\sigma}^\dagger c_{1\sigma}^{\mathstrut}+\text{h.c.}\big)+U\sum_i c_{i\uparrow}^\dagger c_{i\downarrow}^\dagger c_{i\downarrow}^{\mathstrut} c_{i\uparrow}^{\mathstrut}$, where $t$ and $U$ are coefficients and $i\in\{0,1\}$ and $\sigma\in\{\uparrow,\downarrow\}$ are site and spin indices, respectively, so that all index pairs are elements in the one-electron index set $({0\!\uparrow},{0\!\downarrow},{1\!\uparrow},{1\!\downarrow})$.  The parameter space has been restricted by letting $\lambda_n\!\rightarrow i\infty$ for all families $n$ of occupation numbers, except for those associated with two electrons with a zero $z$ component of the total spin.  Permutation symmetry requires that singlet states transform as $A_{1g}$ in the point group $D_{\infty h}$, which means that $\ket{\Psi(\varphi,\phi)}$ is a linear combination of $\ket{\Psi_1}=\big(\!\ket{0110}+\ket{1001}\!\big)/\sqrt{2}$ and $\ket{\Psi_2}=\big(\!\ket{0011}+\ket{1100}\!\big)/\sqrt{2}$.  This requirement is imposed by the symmetry constraints $\lambda_{0110}\!=\!\lambda_{1001}\!=\!\lambda$  and $\lambda_{0011}\!=\!\lambda_{1100}\!=\!-\lambda$.  The parametric equation $\lambda=\lambda(\varphi,\phi)=\phi/2-i\ln\tan\big(\frac{\pi}{4}\!+\!\frac{\varphi}{2}\big)/2$ is defined on the parameter space $\{(\varphi,\phi)\!:\!\varphi\!\in\!(-\pi/2,\pi/2), \phi\!\in\!(-\pi,\pi]\}$, where $\varphi$ and $\phi$ represent the latitude and longitude, respectively, on a sphere.  The energy $E(\varphi,\phi)$ for $t/U=-0.158$ is shown in color in (a) with the colorbar in (b).  The red curves trace the gradient-descent path along $\phi=0$ from the initial guiding state $\ket{\Psi_0}=\big(\!\ket{\Psi_1}+\ket{\Psi_2}\!\big)/\sqrt{2}$ to the ground state at $\varphi\approx1$\,rad.}
	\label{f.2}
\end{figure}

\begin{table*}
	\caption{The coefficients $h_l$, operators $C_{n_l^+}^\dagger C_{n_l^-\vphantom{^+}}\vphantom{^\dagger}$, families $n_l^\pm$, subsets $\dot{\mathcal Q}_l$ and $\vec{\mathcal Q}_l$, and subfamilies $\dot{n}_l^\pm$ and $\vec{n}_l^\pm$ for each interaction in the two-site Hubbard model with the Hamiltonian $\hat H=t\sum_\sigma\big(c_{0\sigma}^\dagger c_{1\sigma}^{\mathstrut}+\text{h.c.}\big)+U\sum_i c_{i\uparrow}^\dagger c_{i\downarrow}^\dagger c_{i\downarrow}^{\mathstrut} c_{i\uparrow}^{\mathstrut}$, where $t$ and $U$ are coefficients and $i\in\{0,1\}$ and $\sigma\in\{\uparrow,\downarrow\}$ are site and spin indices, respectively, so that all index pairs are elements in $\mathcal Q=({0\!\uparrow},{0\!\downarrow},{1\!\uparrow},{1\!\downarrow})$.}
	\centering
	\begin{tabular}{c|c|c|c|c|c|c|c|c|c}
		\hline
		$\quad~~ h_l\quad~~$ & $\quad C_{n_l^+}^\dagger C_{n_l^-\vphantom{^+}}\vphantom{^\dagger}\quad$ & $\qquad n_l^+\qquad$ & $\qquad n_l^-\qquad$ & $\qquad \dot{\mathcal Q}_l\qquad$ & $\qquad \vec{\mathcal Q}_l\qquad$ & $\quad~ {\dot n}_l^+\quad~$  & $\qquad \vec{n}_l^+\qquad$ & $\quad~ {\dot n}_l^-\quad~$  & $\qquad \vec{n}_l^-\qquad$\\
		\hline\hline
		$t$ & $c_{0\uparrow}^\dagger c_{1\uparrow}^{\mathstrut}$ & $(1,0,0,0)$ & $(0,0,1,0)$ & $(0\!\uparrow,1\!\uparrow)$ & $(0\!\downarrow,1\!\downarrow)$ & $(1,0)$ & $(0,0)$ & $(0,1)$ & $(0,0)$\\
		$t$ & $c_{1\uparrow}^\dagger c_{0\uparrow}^{\mathstrut}$ & $(0,0,1,0)$ & $(1,0,0,0)$ & $(0\!\uparrow,1\!\uparrow)$ & $(0\!\downarrow,1\!\downarrow)$ & $(0,1)$ & $(0,0)$ & $(1,0)$ & $(0,0)$\\
		$t$ & $c_{0\downarrow}^\dagger c_{1\downarrow}^{\mathstrut}$ & $(0,1,0,0)$ & $(0,0,0,1)$ & $(0\!\downarrow,1\!\downarrow)$ & $(0\!\uparrow,1\!\uparrow)$ & $(1,0)$ & $(0,0)$ & $(0,1)$ & $(0,0)$\\
		$t$ & $c_{1\downarrow}^\dagger c_{0\downarrow}^{\mathstrut}$ & $(0,0,0,1)$ & $(0,1,0,0)$ & $(0\!\downarrow,1\!\downarrow)$ & $(0\!\uparrow,1\!\uparrow)$ & $(0,1)$ & $(0,0)$ & $(1,0)$ & $(0,0)$\\
		$U$ & $c_{0\uparrow}^\dagger c_{0\downarrow}^\dagger c_{0\downarrow}^{\mathstrut} c_{0\uparrow}^{\mathstrut}$ & $(1,1,0,0)$ & $(1,1,0,0)$ & $(~)$ & $(0\!\uparrow,0\!\downarrow,1\!\uparrow,1\!\downarrow)$ & $(~)$ & $(1,1,0,0)$ & $(~)$ & $(1,1,0,0)$\\
		$U$ & $c_{1\uparrow}^\dagger c_{1\downarrow}^\dagger c_{1\downarrow}^{\mathstrut} c_{1\uparrow}^{\mathstrut}$ & $(0,0,1,1)$ & $(0,0,1,1)$ & $(~)$ & $(0\!\uparrow,0\!\downarrow,1\!\uparrow,1\!\downarrow)$ & $(~)$ & $(0,0,1,1)$ & $(~)$ & $(0,0,1,1)$\\
		\hline
	\end{tabular}
	\label{t.1}
\end{table*}
To make further progress, we need a basis for the Fock space $\mathcal F$.  First, however, we introduce the totally ordered index set $\mathcal{Q}$ for the basis $(\ket{\psi_q})_{q\in\mathcal Q}$ for the one-fermion Hilbert space $\mathcal H$.  The cardinality $Q$ of this set (i.e., the dimension of $\mathcal H$) is herein our measure of the system size.  Each $q\in\mathcal{Q}$ has an occupation number $n_q$ in $\{0,1\}$ that is zero if $\ket{\psi_q}$ is unoccupied and one if it is occupied.  Each family of occupation numbers $n=(n_q)_{q\in\mathcal{Q}}$ in the Cartesian power $\mathcal N=\{0,1\}^Q$ identifies the associated operator
\begin{equation}
	C_n^\dagger = \prod_{q\in\mathcal{Q}}\big(c^{\dagger}_q\big)^{n_q},
	\label{e.6}
\end{equation}
on $\mathcal F$, where $c_q^{\dagger}$ is the fermionic creation operator for each $q$ in $\mathcal Q$.  Using these operators, we generate the Fock states $\ket{n}=C_n^\dagger\ket{0}$, for all $n\in\mathcal N$, and then select the set of all Fock states $\{\ket{n}\}$ to be our basis for $\mathcal F$.

We choose the operator $\hat\lambda(\boldsymbol{\theta})$ be diagonal in this basis, so that
\begin{equation}
	\hat\lambda(\boldsymbol{\theta}) = \sum_{n\in\mathcal N}\lambda_n(\boldsymbol{\theta})\ket{n}\!\bra{n},
	\label{e.7}
\end{equation}
where $\lambda_n(\boldsymbol{\theta})\in\mathbb C$ are complex parametric equations.  Our trial state in Eq.\,(\ref{e.3}) can thus be expressed as
\begin{equation}
 	\ket{\Psi(\boldsymbol{\theta})} = \sum_{n\in\mathcal N}e^{i\lambda_n(\boldsymbol{\theta})}\Psi_{0n}\ket{n},
	\label{e.8}
\end{equation}
where $\Psi_{0n}$ is the component of $\ket{\Psi_0}$ associated with $\ket{n}$.  This form is both general and intuitive.  It is general because if we choose our set of parametric equations $\{\lambda_n(\boldsymbol{\theta})\}$ to be a surjective map of $\Theta$ onto $\mathbb C^{2^Q}$ and our components $\Psi_{0n}$ to be nonzero, for all $n\in\mathcal N$, then the ansatz covers the entire Fock space.  It is intuitive because each $\lambda_n(\boldsymbol{\theta})$ is associated with a Fock state $\ket{n}$, which allows us to both exclude specific Fock states by letting $\lambda_n(\boldsymbol{\theta})\!\rightarrow i\infty$ and impose symmetry constraints provided in terms of Fock states on our set $\{\lambda_n(\boldsymbol{\theta})\}$.  An example of a set of parametric equations that both exclude states and impose symmetry constraints is provided in the caption of Fig.~\ref{f.2} and in Appendix~\ref{A}, for a singlet two-electron trial state of the two-site Hubbard model.  We can even choose our ansatz to depend on number operators as in the recent implementation~\cite{Stenger23} of the Jastrow--Gutzwiller ansatz~\cite{Jastrow55,Gutzwiller63} within the CVQE algorithm, where $\lambda_n(\boldsymbol{\theta})=\sum_{qq'\in\mathcal Q}i\theta_{qq'}n_qn_{q'}$.  Lastly, through the implementation of the guiding state $\ket{\Psi_0}$ on the QPU, we could create ansatzes closer to the ground state than could be achieved by classical methods such as the Hartree--Fock method and variational Monte Carlo.

Because the dimension of $\mathcal F$ increases exponentially with the system size, we cannot generally diagonalize $e^{-i\hat\lambda^\dagger(\boldsymbol{\theta})}\hat He^{i\hat\lambda(\boldsymbol{\theta})}$ on a CPU for large $Q$.  Instead, we note that the expectation value of an operator is a linear function, which allows us to expand the expectation value in Eq.\,(\ref{e.5a}) and diagonalize the operator in each expectation value independently.  Before doing so, however, we express the Hamiltonian, using the operators in Eq.\,(\ref{e.6}), in the form of
\begin{equation}
	\hat H=\sum_{l\in\mathcal{L}}h_l\,C_{n_l^+}^\dagger C_{n_l^-\vphantom{^+}}\vphantom{^\dagger},
	\label{e.9}
\end{equation}
where the set $\mathcal{L}$ contains all indices, for which the coefficients $h_l\in\mathbb C$ are nonzero, and the families $n_l^\pm=(n_{lq}^\pm)_{q\in\mathcal{Q}}$ in $\mathcal N$ are defined by the terms
\begin{subequations}
\begin{align}
	n_{lq}^+=\left\{
	\begin{aligned}
	1,&\qquad\text{if }c_q^\dagger\text{ is \phantom{\underline{not}} present},\\
	0,&\qquad\text{if }c_q^\dagger\text{ is \underline{not} present},
	\end{aligned}
	\right.\\[10 pt]
	n_{lq}^-=\left\{
	\begin{aligned}
	1,&\qquad\text{if }c_q\text{ is \phantom{\underline{not}} present},\\
	0,&\qquad\text{if }c_q\text{ is \underline{not} present}.
	\end{aligned}
	\right.
\end{align}
	\label{e.10}%
\end{subequations}
As an example, the coefficients and families for the two-site Hubbard model have been provided in Table~\ref{t.1}.

We assume that each $l\in\mathcal L$ only affects a subset of the one-fermion states, which we identify by the index set $\dot{\mathcal Q}_l\subset\mathcal Q$, and that the number of these states $\dot{Q}_l$ (which is no more than two for one-fermion interactions, four for two-fermion interactions, etc.) does \underline{not} increase with system size $Q$.  We also define the complementary set $\vec{\mathcal Q}_l\subset\mathcal Q$, which contains the indices of the states that are unaffected by the interaction, and unlike the former, the number of these latter states $\vec{Q_l}=Q-\dot{Q}_l$ \underline{does} increase with system size.  \textit{Note that the dot and arrow accents herein refer to ``a few specific'' and ``all the many other'' indices in $\mathcal Q$, respectively.}  By affected states, we mean states with either an associated creation $c_q^\dagger$ or annihilation operator $c_q$ in the Hamiltonian term---but not both, as a number operator $\hat n_q=c_q^\dagger c_q$ could then be formed, which except for a scalar leaves the states intact.  Thus, the complementary index sets can be expressed as
\begin{equation}
\begin{aligned}
	\dot{\mathcal Q}_l&=(q\in\mathcal Q\,:\,n_{lq}^+\ne n_{lq}^-),\\
	\vec{\mathcal Q}_l&=(q\in\mathcal Q\,:\,n_{lq}^+=n_{lq}^-).
\end{aligned}
	\label{e.11}
\end{equation}
Using these sets, we split the families $n$ with the map
\begin{equation}
	n \mapsto \dot{n}\vec{n},
	\label{e.12}
\end{equation}
into pairs of subfamilies $\dot{n}$ and $\vec{n}$ of $n$, for all $n\in\mathcal N$, where by definition the occupation numbers are matched so that
\begin{equation}
\begin{aligned}
	\dot{n}_q=n_q,&\qquad\text{for all }q\in\dot{\mathcal Q}_l,\\
	\vec{n}_q=n_q,&\qquad\text{for all }q\in\vec{\mathcal Q}_l.
\end{aligned}
	\label{e.13}
\end{equation}
As Appendix~\ref{B} shows, this separation allows us to expand each term in the Hamiltonian in Eq.\,(\ref{e.9}) using a complete set of $2^{\dot{Q}_l}$ hermitian operators, all of which we diagonalize analytically using a set of unitary operators $\{\hat R_{lm}\}$ indexed by $\mathcal M_l=\{x,y\}^{\dot{Q}_l}$.  After this diagonalization, we find that the expectation values in Eq.\,(\ref{e.5}) can be expressed as
\begin{subequations}
\begin{align}
	\Upsilon(\boldsymbol{\theta})=&\sum_{l\in\mathcal L}\sum_{m\in\mathcal M_l}\sum_{n\in\mathcal N}\upsilon_{lmn}e^{-i\lambda_{{\dot n}_l^+\vec{n}}^*(\boldsymbol{\theta})}e^{i\lambda_{{\dot n}_l^-\vec{n}}(\boldsymbol{\theta})}\nonumber\\
	&\times \big|\!\braket{\Psi_0|\hat R_{lm}^\dagger|n}\!\big|^2,\label{e.14a}\\[10 pt]
		\Lambda(\boldsymbol{\theta})=&\sum_{n\in\mathcal N}e^{-2\operatorname{Im}\lambda_{n}(\boldsymbol{\theta})}\big|\!\braket{\Psi_0|n}\!\big|^2,\label{e.14b}
\end{align}
	\label{e.14}%
\end{subequations}
where the complex coefficients are
\begin{equation}
	\upsilon_{lmn} = \frac{\pi_lh_l}{2^{\dot{Q}_l}}N_{ln}V_{lm}Z_{ln},
	\label{e.15}
\end{equation}
where $\pi_l\in\{\pm1\}$ is given by the permutation that separates the fermionic operators indexed by $\dot{\mathcal Q}_l$ and $\vec{\mathcal Q}_l$, and the factors $N_{ln}$, $V_{lm}$, and $Z_{ln}$ are provided in Eq.\,(\ref{b.13}), (\ref{b.18}), and (\ref{b.25}), respectively.  See Appendix~\ref{A} and \ref{B} for an application and derivation of Eq.\,(\ref{e.14}), respectively, and the origin and meaning of each factor in Eq.\,(\ref{e.15}).

As the guiding state $\ket{\Psi_0}$ is prepared on the QPU, we need a map from the Fock space to the Hilbert space of quantum states of the QPU qubit register.  Because qubits are distinguishable, the Hilbert space for a qubit register is a tensor power of the two-dimensional one-qubit Hilbert space $\mathtt H$.  For there to be an isomorphism between the Fock space and this tensor power, we need a register that comprises exactly $Q$ qubits, so that $\dim\mathtt H^{\otimes Q}$ equals $\dim\mathcal F$.  We let $\{\ket{0},\ket{1}\}$ be our basis for each qubit space $\mathtt H$ and define the isomorphism $\mathcal F\rightarrow\mathtt H^{\otimes Q}$ by mapping the Fock states $\ket{n}$ to the tensor products
\begin{equation}
	\ket{n}=\bigotimes_{q\in\mathcal Q}\ket{n_q},
	\label{e.16}
\end{equation}
for all $n\in\mathcal N$.  This Jordan--Wigner mapping~\cite{Ortiz01} transforms the guiding state $\ket{\Psi_0}$ to itself and thus preserves all its components $\Psi_{0n}$.  This makes it straightforward to construct a quantum circuit for $\hat U$ that uses --$X$-- gates to generate the ground state $\ket{n^*}$ of some model of the system within the independent fermion approximation, for which the components are $\Psi_{0n^*}$, and from there introduce weights associated with other Fock states by adding additional gates.

The unitary operators used in the diagonalization are represented by
\begin{equation}
	\hat R_{lm}=\hat\pi_l\bigotimes_{q\in\dot{\mathcal Q}_l}^{\vphantom{}}\hat R_y^{\delta_{m_qx}}\hat R_x^{\delta_{m_qy}}\bigotimes_{q\in\vec{\mathcal Q}_l}^{\vphantom{}}\hat I,
	\label{e.17}
\end{equation}
on $\mathtt H^{\otimes Q}$ for all families $m=(m_q)_{q\in\dot{\mathcal Q}_l}$ in $\mathcal M_l$, where the operator $\hat\pi_l$ is defined such that it permutes the operators on the individual Hilbert spaces $\mathtt H$ in $\mathtt H^{\otimes Q}$ to the order given by $\mathcal Q$ (cf. Appendix~\ref{B}), $\hat R_x$ and $\hat R_y$ are operators that describe one-qubit rotations around the $x$ and $y$ axes by $\pi/2$ and $-\pi/2$, respectively, and $\hat I$ is the identity operator on $\mathtt H$.  The rotation operators $\hat R_x$ and $\hat R_y$ can be implemented in a circuit using the gate sequences --$\sqrt{X}$-- and --$X$--$H$-- (or --$H$--$Z$--), respectively.

In our sampling on the QPU, we use the fact that the probability that a measurement in the basis $\{\ket{n}\}$ for $\mathtt H^{\otimes Q}$ would collapse the state $\hat R\ket{\Psi_0}$, for any unitary operator $\hat R$, to the state $\ket{n}$ associated with a particular outcome $n$ in the sample space $\mathcal N$ is given by the probability mass function
\begin{equation}
	\operatorname{P}[\hat R\Psi_0\!\mapsto\! n] = \big|\!\braket{\Psi_0|\hat R^\dagger|n}\!\big|^2.
	\label{e.18}
\end{equation}
By performing $S$ identical measurements of $\hat R\ket{\Psi_0}$ and recording the outcome $n_s$ of each shot $s$ in some set $\mathcal{S}$, we obtain a collection of families $(n_s^{\hat R})_{s\in\mathcal{S}}$.  Given this sample, we can then apply the law of the unconscious statistician and approximate the expectation value of a function $g(n)$ with the arithmetic mean, which yields
\begin{equation}
	\sum_{n\in\mathcal N}g(n)\operatorname{P}[\hat R\Psi_0\!\mapsto\! n] \approx \frac{1}{S}\sum_{s\in\mathcal{S}}g(n_s^{\hat R}),
	\label{e.19}
\end{equation}
where the sample size $S$ is chosen such that the desired statistical accuracy is attained.

Depending on the particular ansatz of interest---which remarkably could even depend on the sampling itself through the parametric equations---there is not necessarily a unique approach to calculate the expectation values in Eq.\,(\ref{e.14}).  One approach that is guaranteed to work is to collect samples for all the unitary operators $\hat R_{lm}$ (and the identity operator if it has not already been included).  We find that the number of these samples equals the number of nonzero coefficients in the Pauli expansion of the Hamiltonian in the VQE algorithm.  After applying Eq.\,(\ref{e.18}) and (\ref{e.19}), the number of terms in Eq.\,(\ref{e.14}) only increase polynomially with the system size.  Thus, we can then calculate the energy $E(\boldsymbol{\theta})$ in Eq.\,(\ref{e.4}), for any variational parameter vector $\boldsymbol{\theta}$ in $\Theta$, using the CPU.  Consequently, by reusing the collected samples, we perform the optimization entirely on a CPU.  The total number of quantum circuit executions in CVQE is in the most general case given by the number of samples times the number of shots.  As mentioned above, the number of quantum circuit executions in CVQE has as a result, compared to VQE, been reduced by the factor of the number of energy values that needs to be calculated during the optimization process.

Because the energy minimization in the CVQE algorithm is efficient on a CPU, many optimization methods and implementations become available.  One approach is to calculate the energy gradient
\begin{equation}
	\nabla E(\boldsymbol{\theta}) = \frac{\Lambda(\boldsymbol{\theta})\nabla\Upsilon(\boldsymbol{\theta})-\Upsilon(\boldsymbol{\theta})\nabla\Lambda(\boldsymbol{\theta})}{\Lambda^2(\boldsymbol{\theta})},
	\label{e.20}
\end{equation}
using the gradients
\begin{subequations}
\begin{align}
	\nabla\Upsilon(\boldsymbol{\theta}) =& \sum_{l\in\mathcal L}\sum_{m\in\mathcal M_l}\sum_{n\in\mathcal N}\upsilon_{lmn}e^{-i\lambda_{{\dot n}_l^+\vec{n}}^*(\boldsymbol{\theta})}e^{i\lambda_{{\dot n}_l^-\vec{n}}(\boldsymbol{\theta})}\nonumber\\
	&\times\big[-i\nabla\lambda_{{\dot n}_l^+\vec{n}}^*(\boldsymbol{\theta})+i\nabla\lambda_{{\dot n}_l^-\vec{n}}(\boldsymbol{\theta})\big]\big|\!\braket{\Psi_0|\hat R_{lm}^\dagger|n}\!\big|^2,\\
	\nabla\Lambda(\boldsymbol{\theta}) =& \sum_{n\in\mathcal N}e^{-2\operatorname{Im}\lambda_{n}(\boldsymbol{\theta})}\big[-2\nabla\operatorname{Im}\lambda_{n}(\boldsymbol{\theta})\big]\big|\!\braket{\Psi_0|n}\!\big|^2,
\end{align}
	\label{e.21}%
\end{subequations}
using the collected samples and analytical derivatives for the gradients of the parametric equations.  The energy gradient, along with higher derivatives, if needed, can be used in any iterative optimization method in the form of
\begin{equation}
	\boldsymbol{\theta}_{k+1}=\boldsymbol{\theta}_k+f[\nabla E(\boldsymbol{\theta}_k),...],
	\label{e.22}
\end{equation}
where each $k=0,1,2,...$ successively generates a new parameter vector, starting from the initial trial vector $\boldsymbol{\theta}_0$, and $f[\nabla E(\boldsymbol{\theta}_k),...]$ is a function that defines the method.  One method of this form is gradient descent $f[\nabla E(\boldsymbol{\theta}_k),...]=-\gamma_k\nabla E(\boldsymbol{\theta}_k)$, which we used for the optimization in Fig.\,\ref{f.2} with the step size $\gamma_k=1$ (For more details, see Appendix~\ref{A}).  If converged, the solution vector $\boldsymbol{\theta}^*$ minimizes $E(\boldsymbol{\theta})$, and the energy $E(\boldsymbol{\theta}^*)$ is the sought upper bound for the ground-state energy.

This work has been supported by the Office of Naval Research (ONR) through the U.S. Naval Research Laboratory (NRL).  J.P.T.S. thanks the National Research Council Research Associateship Programs for support during his postdoctoral tenure at NRL.

\appendix
\renewcommand{\appendixname}{APPENDIX}
\renewcommand{\thesection}{\Alph{section}}

\section{}
\label{A}

To test the closed-form expression of the energy $E(\boldsymbol{\theta})$ in Eq.\,(\ref{e.4}) given by Eq.\,(\ref{e.14}), let us consider an electronic system described by a two-site Hubbard model, for which we can obtain the energy directly by calculating the expectation value analytically for all $\boldsymbol{\theta}$ in the parameter space $\Theta$.

If we denote the site $i\in\{0,1\}$ and the spins $\sigma\in\{\uparrow,\downarrow\}$, the Hamiltonian for this system can then be expressed as
\begin{equation}
	\hat H=t\sum_\sigma\big(c_{0\sigma}^\dagger c_{1\sigma}+\text{h.c.}\big)+U\sum_i \hat n_{i\uparrow}\hat n_{i\downarrow},
	\label{a.1}
\end{equation}
where $t$ and $U$ are the model hopping and Hubbard-U parameters, respectively.  Following the approach in the main text, we introduce the basis $\{\ket{\psi_{i\sigma}}\}$ for the 4-dimensional, one-electron Hilbert space formed by the four states $\ket{\psi_{i\sigma}}=c_{i\sigma}^\dagger\ket{0}$ indexed by the site-spin set $\mathcal Q=\{0\!\!\uparrow,0\!\!\downarrow,1\!\!\uparrow,1\!\!\downarrow\}$.  We construct the basis $\{\ket{n}\}$ for the Fock space $\mathcal F$ from the Fock states $\ket{n}=C_n^\dagger\ket{0}$, which are labeled by the families $n=(n_{i\sigma})$, where $n_{i\sigma}$ is the number of electrons in $\ket{\psi_{i\sigma}}$.  From Eq.\,(\ref{e.6}), which in this example is given by
\begin{equation}
	C_n^\dagger = \prod_{i\sigma}\big(c^{\dagger}_{i\sigma}\big)^{n_{i\sigma}},
	\label{a.2}
\end{equation}
we find that the Fock state $\ket{1001}=c^{\dagger}_{0\uparrow}c^{\dagger}_{1\downarrow}\ket{0}$, say, can be identified by the family that has one electron in $\ket{\psi_{0\uparrow}}$, zero electrons in $\ket{\psi_{0\downarrow}}$, zero electrons in $\ket{\psi_{1\uparrow}}$, and one electron in $\ket{\psi_{1\downarrow}}$.

For our test demonstration, we choose our guiding state $\ket{\Psi_0}$ in Eq.\,(\ref{e.2}) to be defined by
\begin{equation}
	\hat U=\bigotimes_{q\in\mathcal Q}\frac{\ket{0}\!\bra{0}+\ket{0}\!\bra{1}+\ket{1}\!\bra{0}-\ket{1}\!\bra{1}}{\sqrt{2}},
	\label{a.3}
\end{equation}
so that we can calculate the probability mass functions in Eq.\,(\ref{e.18}) analytically.  This allows us to also calculate the energy using Eq.\,(\ref{e.4}) and (\ref{e.14}) analytically and verify the result against the expectation value of the Hamiltonian in the trial state $\ket{\Psi(\boldsymbol{\theta})}$.  If one instead were to perform the sampling of the guiding state
\begin{equation}
	\ket{\Psi_0}=\frac{1}{4}\sum_{n\in\mathcal N}\ket{n}
	\label{a.4}
\end{equation}
on the QPU following the CVQE algorithm as illustrated in Fig.\,\ref{f.1}, the operator $\hat U$ would be implemented by a quantum circuit that executes a Hadamard gate --$H$--, for each qubit in the register.  

We are specifically interested in the two-electron, spin-singlet ground state.  In our basis $\{\ket{n}\}$ for $\mathcal F$, there are six two-electron Fock states, four of which with the $z$ component of the total spin being zero.  These states are $\ket{0011}$, $\ket{0110}$, $\ket{1001}$, and $\ket{1100}$.  We also note that the two-site Hubbard model has point group symmetry $D_{\infty h}$ and the relevant symmetry operation with respect to the mentioned four Fock states is the inversion operator.  Our goal is to form symmetry-adapted states, which are either symmetric or antisymmetric under inversion.  Because electrons are fermions, the states must be antisymmetric with respect to the exchange of the two electrons.  As spin-singlet states are antisymmetric under this exchange, our spatial symmetry-adapted states must be symmetric with respect to the electron exchange.  This requires that the ground state transforms as the irreducible representation $A_{1g}$ of $D_{\infty h}$.  There are two such symmetry-adapted states that can be formed by the basis states $\ket{0011}, \ket{0110}$, $\ket{1001}$, and $\ket{1100}$.  They are
\begin{subequations}
\begin{align}
	\ket{\Psi_1}&=\frac{1}{\sqrt{2}}\Big(\ket{0110}+\ket{1001}\Big),\\
	\ket{\Psi_2}&=\frac{1}{\sqrt{2}}\Big(\ket{0011}+\ket{1100}\Big).
\end{align}
	\label{a.5}%
\end{subequations}
It can easily be verified that both $\ket{\Psi_1}$ and $\ket{\Psi_2}$ are symmetric under inversion, which in the two-site Hubbard model exchanges the sites $0$ and $1$.  To impose this required symmetry, we choose $\lambda_{0110}=\lambda_{1001}=\lambda$ and $\lambda_{0011}=\lambda_{1100}=-\lambda$, and $\lambda_n\rightarrow i\infty$ for all families $n$ not in $\{0011,0110,1001,1100\}$, where
\begin{equation}
	\lambda(\varphi,\phi)=\frac{\phi}{2}-\frac{i}{2}\ln\tan\big(\frac{\pi}{4}\!+\!\frac{\varphi}{2}\big),
	\label{a.6}
\end{equation}
is defined on the parameter space $\Theta=\{(\varphi,\phi)\!:\!\varphi\!\in\!(-\pi/2,\pi/2), \phi\!\in\!(-\pi,\pi]\}$.  This form of $\lambda(\varphi,\phi)$ conveniently represents the 
normalized trial state
\begin{equation}
	\ket{\Psi(\varphi,\phi)}=\sin\big(\frac{\pi}{4}\!+\!\frac{\varphi}{2}\big)e^{i\phi/2}\ket{\Psi_1}+\cos\big(\frac{\pi}{4}\!+\!\frac{\varphi}{2}\big)e^{-i\phi/2}\ket{\Psi_2}
	\label{a.7}
\end{equation}
on the Bloch sphere, where $\varphi$ and $\phi$ are the latitude and longitude, respectively, and $\ket{\Psi_1}$ and $\ket{\Psi_2}$ are the north and south pole, respectively (cf. Fig.~\ref{f.2}).

For our choice of $\hat U$, the probability mass function for a measurement of the guiding state $\ket{\Psi_0}$ is
\begin{equation}
	\operatorname{P}[\Psi_0\!\mapsto\! n]=\frac{1}{16}.
	\label{a.8}
\end{equation}
After some algebra, one finds that the energy denominator in Eq.\,(\ref{e.14b}) for this model is
\begin{equation}
	\Lambda(\varphi,\phi) = \frac{1}{4\cos\varphi}.
	\label{a.9}
\end{equation}

Before we can calculate the corresponding energy numerator in Eq.\,(\ref{e.14a}), we need to identify the coefficients $h_l$ and families $n_l^\pm$ in Eq.\,(\ref{e.9}) for the Hamiltonian in Eq.\,(\ref{a.1}).  This is straightforward as the coefficients $h_l$ are directly given and the families $n_l^+$ and $n_l^-$ merely identify which index pairs $i\sigma\in\{0\!\!\uparrow,0\!\!\downarrow,1\!\!\uparrow,1\!\!\downarrow\}$ have creation and annihilation operators, respectively, in the term $l$.  For example, the first term in Eq.\,(\ref{a.1}), $t\,c_{0\uparrow}^\dagger c_{1\uparrow}$, corresponds to $h_1=t$, $n_1^+=(1,0,0,0)$, and $n_1^-=(0,0,1,0)$ in Eq.\,(\ref{e.9}).  For a full list of the identified coefficients and families, see Table~\ref{t.1}.

For each interaction $l\in\mathcal L$, we also need to identify the affected index pairs $i\sigma\in\mathcal Q$, which form the subset $\dot{\mathcal Q}_l$.  This and its complementary subset are given by Eq.\,(\ref{e.11}) and also provided in Table~\ref{t.1}.  Note that the index pairs for number operators are not considered affected, and therefore, the subset $\dot{\mathcal Q}_l$ is empty for the two-electron terms.  Once the affected and unaffected subsets have been defined, we split the family of occupation numbers in $n_l^\pm$ into a collection of occupation numbers affected $\dot n_l^\pm$ and unaffected $\vec n_l^\pm$ by each interaction, in accordance with Eq.\,(\ref{e.12}) and (\ref{e.13}).  The resulting collections are also provided in Table~\ref{t.1}.

Next, we need to identify the coefficients in Eq.\,(\ref{e.15}).  The sign $\pi_l$ can be negative only for interactions that contain creation, annihilation, and number operators; permutations of creation or annihilation operators ordered by Eq.\,(\ref{e.6}) may then be necessary for the number operators to form.  As this is not the case for any of the interactions in the Hubbard model, we have $\pi_l=+1$ for all $l\in\mathcal L$.

The coefficients $N_{l{n}}$ in Eq.\,(\ref{b.13}) incorporate a Kronecker delta function for each number operator in the interaction, which ensures that an electron occupies each spin-orbital that has a number operator.  These coefficients are unity for all one-electron terms in the Hamiltonian and $\delta_{n_{i\uparrow}1}\delta_{n_{i\downarrow}1}$ for the two-electron terms identified by the site $i$.  Thus, only when there are two electrons on site $i$ does the corresponding two-electron term contribute to the energy.

\begin{table}[h]
	\caption{Applicable expansion coefficients $V_{lm}$ for each interaction identified by $C_{n_l^+}^\dagger C_{n_l^-\vphantom{^+}}\vphantom{^\dagger}$ and the expansion index $m$.}
	\centering
	\begin{tabular}{c|c|c|c|c|c}
		\hline
		$\quad C_{n_l^+}^\dagger C_{n_l^-\vphantom{^+}}\vphantom{^\dagger}\quad$ & $\quad V_{l()}\quad$ & $~ V_{l(x,x)}~$  & $~ V_{l(x,y)}~$ & $~ V_{l(y,x)}~$  & $~ V_{l(y,y)}~$\\
		\hline\hline
		$c_{0\uparrow}^\dagger c_{1\uparrow}^{\mathstrut}$ &  & $+1$ & $+i$ & $-i$ & $+1$\\
		$c_{1\uparrow}^\dagger c_{0\uparrow}^{\mathstrut}$ &  & $+1$ & $-i$ & $+i$ & $+1$\\
		$c_{0\downarrow}^\dagger c_{1\downarrow}^{\mathstrut}$ &  & $+1$ & $+i$ & $-i$ & $+1$\\
		$c_{1\downarrow}^\dagger c_{0\downarrow}^{\mathstrut}$ &  & $+1$ & $-i$ & $+i$ & $+1$\\
		$c_{0\uparrow}^\dagger c_{0\downarrow}^\dagger c_{0\downarrow}^{\mathstrut} c_{0\uparrow}^{\mathstrut}$ & $+1$ &  &  &  & \\
		$c_{1\uparrow}^\dagger c_{1\downarrow}^\dagger c_{1\downarrow}^{\mathstrut} c_{1\uparrow}^{\mathstrut}$ & $+1$ &  &  &  & \\
		\hline
	\end{tabular}
	\label{t.a1}
\end{table}
The coefficients $V_{lm}$ in Eq.\,(\ref{b.18}) contain the phase factors that result from the expansion of the interaction terms represented on the qubit register Hilbert space.  They are listed for each interaction in the Hubbard model in Table~\ref{t.a1}.

To determinate the probability mass function for measurements of the state $\hat R_{lm}\ket{\Psi_0}$, we first note that the guiding state in Eq.\,(\ref{a.4}) is represented by the tensor power
\begin{equation}
	\ket{\Psi_0}=\ket{+}^{\otimes Q},
	\label{a.10}
\end{equation}
on $\mathtt H^{\otimes Q}$, where $\ket{+}=(\ket{0}+\ket{1})/\sqrt{2}$.  From the maps
\begin{subequations}
\begin{align}
	\hat R_x\ket{+}=&e^{-i\pi/4}\ket{+},\\
	\hat R_y\ket{+}=&\ket{0},
\end{align}
	\label{a.11}%
\end{subequations}
then follow the probability mass function
\begin{equation}
	\operatorname{P}[\hat R_{lm}\Psi_0\!\mapsto\! n]=\frac{1}{2^{\vec Q_l}}\prod_{q\in\dot{\mathcal Q}_l}\Big[\delta_{n_q0}^{\delta_{m_qx}}+2^{-\delta_{m_qy}}\Big].
	\label{a.12}
\end{equation}

Using this probability mass function, the numerator in Eq.\,(\ref{e.14a}) can be expressed as
\begin{equation}
	\Upsilon(\varphi,\phi) = \frac{t}{2}\cos\big[2\operatorname{Re}\lambda(\varphi,\phi)\big]+\frac{U}{8}e^{2\operatorname{Im}\lambda(\varphi,\phi)},
	\label{a.13}
\end{equation}
where we have used
\begin{equation}
	\frac{1}{2^{\dot{Q}_l}}\sum_{m\in\mathcal M_l}\sum_{\dot n\in\dot{\mathcal N}}V_{lm}Z_{ln}\operatorname{P}[\hat R_{lm}\Psi_0\!\mapsto\! n]=\frac{1}{16},
	\label{a.14}
\end{equation}
for all interactions $l\in\mathcal L$, where $\dot{\mathcal N}=\{0,1\}^{\dot{Q}_l}$.  Inserting Eq.\,(\ref{a.6}) into Eq.\,(\ref{a.13}) and dividing by Eq.\,(\ref{a.9}), finally yields the energy 
\begin{equation}
	E(\varphi,\phi)=2t\cos\varphi\cos\phi+\frac{U}{2}\big(1-\sin\varphi\big).
	\label{a.15}
\end{equation}

To verify the above result, we find the Hamiltonian matrix elements
\begin{equation}
\begin{split}
	\braket{\Psi_1|\hat H|\Psi_1}&=0,\\
	\braket{\Psi_1|\hat H|\Psi_2}&=2t,\\
	\braket{\Psi_2|\hat H|\Psi_1}&=2t,\\
	\braket{\Psi_2|\hat H|\Psi_2}&=U,
\end{split}
	\label{a.16}%
\end{equation}
and apply these to calculate the expectation value
\begin{equation}
	E(\varphi,\phi) = \braket{\Psi(\varphi,\phi)|\hat H|\Psi(\varphi,\phi)}.
	\label{a.17}
\end{equation}
of the Hamiltonian in the trial state in Eq.\,(\ref{a.7}).  As expected, we find the same expression for the energy shown in Eq.\,(\ref{a.15}).  Note that this direct calculation is of course not generally available as the dimension of the space that the trial state and Hamiltonian is represented on increases exponentially with the system size.  As shown herein, however, the energy in Eq.\,(\ref{e.4}) can still be obtained with CPU resources that only increases polynomially with system size by calculating Eq.\,(\ref{e.14}), provided that measurements samples have first been collected on the QPU so that the sample mean in Eq.\,(\ref{e.19}) can be applied.

\begin{table}[h]
	\caption{Optimization of the parameter $\varphi$ in the ansatz for the two-site Hubbard model ($t/U=-0.158$) and the associated energy $E(\varphi_k,0)$.}
	\centering
	\begin{tabular}{c|c}
		\hline
		$\qquad~~\varphi_k~[\text{deg}]\qquad~~$ & $~~ E(\varphi_k,0)~[\text{units of }U]~~$\\
		\hline\hline
        0 	&	0.1840\\
   35.1077	&	-0.0461\\
   47.5575	&	-0.0822\\
   53.0757	&	-0.0896\\
   55.5872	&	-0.0911\\
   56.7362	&	-0.0914\\
   57.2624	&	-0.0915\\
   57.5034	& 	-0.0915\\
   57.6138	&	-0.0915\\
   57.6644	&	-0.0915\\
   57.6875	&	-0.0915\\
   57.6981	&	-0.0915\\
   57.7030	&	-0.0915\\
   57.7052	&	-0.0915\\
   57.7063	&	-0.0915\\
   57.7067	&	-0.0915\\
   57.7069	&	-0.0915\\
   57.7070	&	-0.0915\\
   57.7071	&	-0.0915\\
   57.7071	&	-0.0915\\
		\hline
	\end{tabular}
	\label{t.a2}
\end{table}
The gradient of the energy is
\begin{equation}
	\nabla E(\varphi,\phi) = -\bigg[2t\sin\varphi\cos\phi+\frac{U}{2}\cos\varphi\bigg]\boldsymbol{e}_\varphi-2t\sin\phi\,\boldsymbol{e}_\phi,
	\label{a.18}
\end{equation}
where $\boldsymbol{e}_\varphi$ and $\boldsymbol{e}_\phi$ are the standard basis vectors for a spherical coordinate system with a constant radius.  Using the same basis vectors, the parameter vector is
\begin{equation}
	\boldsymbol{\theta}=\varphi\,\boldsymbol{e}_\varphi+\phi\,\boldsymbol{e}_\phi.
	\label{a.19}
\end{equation}
Starting from the initial trial vector $\boldsymbol{\theta}_0=\boldsymbol{0}$, the new parameter vectors in gradient descent are given by
\begin{equation}
	\boldsymbol{\theta}_{k+1}=\boldsymbol{\theta}_k-\nabla E(\boldsymbol{\theta}_k),
	\label{a.20}
\end{equation}
for $k=0,1,2,...$, where we have chosen the step-size parameter $\gamma_k=1$ in Eq.\,(\ref{e.22}).  In coordinate form, we have
\begin{subequations}
\begin{align}
	\varphi_{k+1}&=\varphi_k+2t\sin\varphi_k\cos\phi_k+\frac{U}{2}\cos\varphi_k,\label{a.21a}\\
	\phi_{k+1}&=\phi_k+2t\sin\phi_k.\label{a.21b}
\end{align}
	\label{a.21}%
\end{subequations}
As $\phi_0=0$, we find from the latter equation that $\phi_k=0$, for all $k$.  For negative $t$, we find that $\partial^2E/\partial\phi^2(\varphi,0)>0$, for $\varphi\in(-\pi/2,\pi/2)$, and thus $\phi=0$ is a minimum in the direction $\boldsymbol{e}_\phi$.  The optimization  in the direction $\boldsymbol{e}_\varphi$ is given by Eq.\,(\ref{a.21a}) with $\cos\phi_k=1$.  The first 20 parameters $\varphi_k$ are shown in Table~\ref{t.a2}.  The minimized solution is $(\varphi^*,\phi^*)\approx(57.7071^\circ,0)$ and the associated minimized energy $E(\varphi^*,\phi^*)\approx-0.0915U$.

\section{}
\label{B}

A critical component of the CVQE algorithm is the closed-form expression for the energy $E(\boldsymbol{\theta})$ in Eq.\,(\ref{e.4}) given by Eq.\,(\ref{e.14}) that can be calculated efficiently on the CPU for any parameter vector $\boldsymbol{\theta}$ in the parameter space $\Theta$ using Eq.\,(\ref{e.19}) with the measurement samples collected priorly on the QPU.  Below, we provide more details on how Eq.\,(\ref{e.14}) was derived from the expectation values in Eq.\,(\ref{e.5}).

To calculate expectation values with the assistance of a QPU without having to introduce extra ancillary qubits, the operators in the expectation values must be diagonal in the measurement basis.  The challenge is that diagonalizing the operators in Eq.\,(\ref{e.5}) analytically or numerically is generally hard as the dimension of the Fock space $\dim\mathcal F=2^Q$ increases exponentially with the system size $Q$.  Fortunately, however, the expectation value of an operator is a linear function.  Thus, if we consider the system of interest being a collection of interactions indexed by $\mathcal L$ and described by the Hamiltonians
\begin{equation}
	\hat H_l=h_l\,C_{n_l^+}^\dagger C_{n_l^-\vphantom{^+}}\vphantom{^\dagger},
	\label{b.1}
\end{equation}
on $\mathcal F$, for all $l\in\mathcal L$, the expectation value in Eq.\,(\ref{e.5a}) is the linear combination
\begin{equation}
	\braket{\Psi_0|e^{-i\hat\lambda^\dagger(\boldsymbol{\theta})}\hat He^{i\hat\lambda(\boldsymbol{\theta})}|\Psi_0}=\sum_{l\in\mathcal L}\braket{\Psi_0|e^{-i\hat\lambda^\dagger(\boldsymbol{\theta})}\hat H_le^{i\hat\lambda(\boldsymbol{\theta})}|\Psi_0}
	\label{b.2}
\end{equation}
over the individual interactions.

The order of the creation and annihilation operators in $\hat H_l$ imposed by index set $\mathcal Q$ via Eq.\,(\ref{e.6}) is generally fine, except for those interactions with some---but not all---creation and annihilation operators forming number operators.  In that case, the creation and annihilation operators might need to be reordered to allow for all possible number operators to form.  One order that always works is given by the permutation $\hat\pi_l$ defined such that the mapping $\hat\pi_l:\mathcal Q\rightarrow\mathcal Q$ produces
\begin{equation}
	\hat\pi_l\mathcal Q\mapsto\dot{\mathcal Q}_l\frown\vec{\mathcal Q}_l,
	\label{b.3}
\end{equation}
where $\frown$ refers to concatenation (i.e., the order of indices in $\mathcal Q$ is preserved except that all indices $q\in\dot{\mathcal Q}_l$ have been moved to the left of all $q\in\vec{\mathcal Q}_l$).  To obtain this order among our creation and annihilation operators, we apply the inverse permutation $\hat\pi_l^{-1}$ to each instance of the operator in Eq.\,(\ref{e.6}), so that
\begin{equation}
	\hat\pi_l^{-1}C_n^\dagger = \Big[\prod_{q\in\hat\pi_l\mathcal{Q}}\big(c^{\dagger}_q\big)^{n_q}\Big],\\
	\label{b.4}
\end{equation}
for all $n\in\mathcal N$.  From the fermionic anti-commutation relation $\{c_q,c_{q'}\}=0$, for all $qq'\in\mathcal Q$, it then follows that
\begin{equation}
	\hat H_l=\pi_lh_l\Big[\prod_{q\in\hat\pi_l\mathcal{Q}}\big(c^{\dagger}_q\big)^{n_{lq}^+}\Big]\Big[\prod_{q\in\hat\pi_l\mathcal{Q}}\big(c^{\dagger}_q\big)^{n_{lq}^-}\Big]^\dagger,
	\label{b.5}
\end{equation}
for all $l\in\mathcal L$, where the sign $\pi_l\in\{\pm1\}$ depends on the families $n_l^\pm$ that, along with the coefficient $h_l\in\mathbb C$, specify the interaction as described in the main text.  The order of the creation and annihilation operators given by the permutation $\hat\pi_l$ allows number operators $\hat n_q=c^{\dagger}_qc_q$, for all $q\in\mathcal Q$, to form at the interface between the two products in Eq.\,(\ref{b.5}).  As $[\hat n_q,c_{q'}^\dagger]=[\hat n_q,c_{q'}]=0$, for all $qq'\in\mathcal Q\,:\,q\ne q'$, we find that
\begin{equation}
	\hat H_l = \pi_lh_l\,\hat C_l\hat N_l,
	\label{b.6}
\end{equation}
where
\begin{subequations}
\begin{align}
	\hat C_l&=\Big[\prod_{q\in\dot{\mathcal Q}_l}\big(c^{\dagger}_q\big)^{n_{lq}^+}\Big]\Big[\prod_{q\in\dot{\mathcal Q}_l}\big(c^{\dagger}_q\big)^{n_{lq}^-}\Big]^\dagger,\\
	\hat N_l&=\prod_{q\in\vec{\mathcal Q}_l}\hat n_q^{n_{lq}^+}.
\end{align}
	\label{b.7}%
\end{subequations}

Before proceeding, let us turn to the other operator $e^{i\hat\lambda(\boldsymbol{\theta})}$ in Eq.\,(\ref{b.2}).  Defining the projection operator $\hat P_n=\ket{n}\!\bra{n}$ and using the property $\hat P_n\hat P_{n'}=\delta_{nn'}\hat P_n$, for all $n,n'\in\mathcal N$, one finds from the Taylor expansions of $e^{i\hat\lambda(\boldsymbol{\theta})}$ and $e^{i\lambda_n(\boldsymbol{\theta})}$ that
\begin{equation}
	e^{i\hat\lambda(\boldsymbol{\theta})} = \sum_{n\in\mathcal N}e^{i\lambda_n(\boldsymbol{\theta})}\hat P_n.
	\label{b.8}
\end{equation}
Because $\hat P_n$ is diagonal in the basis $\{\ket{n}\}$, it can be represented by a product of number operators.  Moreover, because these number operators commute, we can put them in any order, including the order given by the permutation $\hat\pi_l$.  Using the map in Eq.\,(\ref{e.12}) that is associated with the perturbation map in Eq.\,(\ref{b.3}), this order yields
\begin{equation}
	e^{i\hat\lambda(\boldsymbol{\theta})} = \sum_{\dot{n}\vec{n}}e^{i\lambda_{\dot{n}\vec{n}}(\boldsymbol{\theta})}\hat P_{\dot{n}}\hat P_{\vec{n}},
	\label{b.9}
\end{equation}
where the two projection operator factors can be expressed as
\begin{subequations}
\begin{align}
	\hat P_{\dot{n}}&=\prod_{q\in\dot{\mathcal Q}_l}\hat n_q^{\dot n_q}(1-\hat n_q)^{(1-\dot n_q)},\\
	\hat P_{\vec{n}}&=\prod_{q\in\vec{\mathcal Q}_l}\hat n_q^{\vec n_q}(1-\hat n_q)^{(1-\vec n_q)}.
\end{align}
	\label{b.10}
\end{subequations}
Again after applying Eq.\,(\ref{e.12}) and the commutation property of number operators, we can write the operator products in Eq.\,({\ref{b.2}) as
\begin{align}
	e^{-i\hat\lambda^\dagger(\boldsymbol{\theta})}\hat H_le^{i\hat\lambda(\boldsymbol{\theta})} = \pi_lh_l\sum_{\dot{n}\vec{n}\dot{n}'\vec{n}'}&e^{-i\lambda_{\dot{n}\vec{n}}^*(\boldsymbol{\theta})}e^{i\lambda_{\dot{n}'\vec{n}'}(\boldsymbol{\theta})}\nonumber\\
	\times&\hat P_{\dot{n}}\hat C_l\hat P_{\dot{n}'}\hat P_{\vec{n}}\hat N_l\hat P_{\vec{n}'}
	\label{b.11}
\end{align}
for all $l\in\mathcal L$.  From the fermionic anti-commutation relation $\{c_q,c_{q'}^\dagger\}=\delta_{qq'}$, for all $qq'\in\mathcal Q$, follows
\begin{subequations}
\begin{align}
	\hat P_{\dot{n}}\hat C_l\hat P_{\dot{n}'}=&\hat C_l\delta_{\dot n\dot{n}_l^+}\delta_{\dot n'\dot{n}_l^-},\\
	\hat P_{\vec{n}}\hat N_l\hat P_{\vec{n}'}=&N_{l{n}}\hat P_{\vec{n}}\delta_{\vec{n}\vec{n}'},
\end{align}
	\label{b.12}%
\end{subequations}
where
\begin{equation}
	N_{l{n}} =\prod_{q\in\vec{\mathcal Q}_l}\delta_{n_q1}^{n_{lq}^+}
	\label{b.13}
\end{equation}
is the eigenvalue of the operator $\hat N_l$ for the state $\ket{n}$, which is one in the case $n_q=1$ for every $q\in\vec{\mathcal Q}_l$ that has a number operator in $\hat N_l$, and zero otherwise.  Inserting Eq.\,(\ref{b.12}), the operator in Eq.\,(\ref{b.11}) becomes
\begin{equation}
	e^{-i\hat\lambda^\dagger(\boldsymbol{\theta})}\hat H_le^{i\hat\lambda(\boldsymbol{\theta})} = \pi_lh_l\sum_{\vec n}N_{ln}e^{-i\lambda_{\dot{n}_l^+\vec n}^*(\boldsymbol{\theta})}e^{i\lambda_{\dot{n}_l^-\vec n}(\boldsymbol{\theta})}\hat C_l\hat P_{\vec n}.
	\label{b.14}
\end{equation}

To achieve the shallowest possible measurement circuits on the QPU, we want to represent the operators $\hat C_l$ and $\hat P_{\vec n}$ on the Hilbert space $\mathtt H^{\otimes Q}$ for the qubit register.  As the guiding state $\ket{\Psi_0}$ implemented on this space is the same for all interactions, we use the global index set $\mathcal Q$ to fix the order of the individual qubit spaces $\mathtt H$ in the register space $\mathtt H^{\otimes Q}$.  The downside with this fixed order $\mathcal Q$, however, is that it does not separate the spaces that contain states that are affected and unaffected by each interaction.  To circumvent this shortcoming, we work with operators on $\mathtt H^{\otimes Q}$ that are ordered by $\hat\pi_l\mathcal Q$ and apply the permutation operator $\hat\pi_l$ defined in Eq.\,(\ref{b.3}) to rearrange the individual qubit operators and restore the fixed order $\mathcal Q$.

The isomorphism $\mathcal F\rightarrow\mathtt H^{\otimes Q}$ given by Eq.\,(\ref{e.16}) is consistent with the Jordan--Wigner transformation~\cite{Jordan28}, which is represented on $\mathtt H^{\otimes Q}$ by
\begin{equation}
	c_q^\dagger = \hat\pi_l\bigotimes_{\substack{q'\in\hat\pi_l\mathcal Q \\ q'<q}}\hat\sigma_z\bigotimes\frac{\hat\sigma_x-i\hat\sigma_y}{2}\bigotimes_{\substack{q'\in\hat\pi_l\mathcal Q \\ q'>q}}\hat\sigma_0,
	\label{b.15}
\end{equation}
for all $q\in\mathcal Q$, where the Pauli operators $\hat\sigma_x$, $\hat\sigma_y$, and $\hat\sigma_z$ are represented by the associated Pauli matrices, and the identity operator $\hat\sigma_0$ by the identity matrix, when the basis states $\ket{0}$ and $\ket{1}$ for $\mathtt H$ are mapped to the column vectors $(1,0)^\intercal$ and $(0,1)^\intercal$ for the vector space $\mathbb C^2$, respectively.  As the binary relation $<$ has been defined with respect to the elements in $\pi_l\mathcal Q$ rather than $\mathcal Q$, the string operator depends on the interactions.  While this approach might appear to unnecessarily complicate matters, the advantage is that the permutation in Eq.\,(\ref{b.3}) was specifically chosen such that the representation of the operators on the right-hand side of Eq.\,(\ref{b.14}) on $\mathtt H^{\otimes Q}$,
\begin{subequations}
\begin{align}
	\hat C_l=&\hat\pi_l\bigotimes_{q\in\dot{\mathcal Q}_l}\bigg(\frac{\hat\sigma_x-i\hat\sigma_y}{2}\bigg)^{n_{lq}^+}\bigg(\frac{\hat\sigma_x+i\hat\sigma_y}{2}\bigg)^{n_{lq}^-}\bigotimes_{q\in\vec{\mathcal Q}_l}\hat\sigma_0,\label{b.16a}\\
	\hat P_{\vec{n}}=& \hat\pi_l\bigotimes_{q\in\dot{\mathcal Q}_l}\hat\sigma_0\bigotimes_{q\in\vec{\mathcal Q}_l}\frac{\hat\sigma_0+(-1)^{\vec n_q}\hat\sigma_z}{2}\label{b.16b},
\end{align}
	\label{b.16}%
\end{subequations}
is unaffected by the string operator.  Consequently, we do not need to track the state-dependent sign that normally results from the $\hat\sigma_z$ operators in the string operator.

Rather than diagonalizing Eq.\,(\ref{b.16a}) directly, which would lead to unnecessarily large measurement circuits, we expand each operator on $\mathtt H$ in the basis $\{\hat\sigma_0,\hat\sigma_x,\hat\sigma_y,\hat\sigma_z\}$.  This expansion yields
\begin{equation}
	\hat C_l=\frac{1}{2^{\dot{Q}_l}}\sum_{m\in\mathcal M_l}V_{lm}\hat V_{lm},
	\label{b.17}
\end{equation}
where $m=(m_q)_{q\in\dot{\mathcal Q}_l}$ are indexed families obtained from the Cartesian power $\mathcal M_l=\{x,y\}^{\dot{Q}_l}$,
\begin{equation}
	V_{lm} =\prod_{q\in\dot{\mathcal Q}_l}\Big[(-1)^{n_{lq}^+}i\Big]^{\delta_{m_qy}}
	\label{b.18}
\end{equation}
are expansion coefficients, and
\begin{equation}
	\hat V_{lm}=\hat\pi_l\bigotimes_{q\in\dot{\mathcal Q}_l}^{\vphantom{}}\hat\sigma_{m_q}\bigotimes_{q\in\vec{\mathcal Q}_l}\hat\sigma_0,
	\label{b.19}
\end{equation}
are hermitian operators.  The purpose of $\delta_{m_qy}$ is to pick up the phase factor $\mp i$, if and only if $m_q=y$, where the sign is minus (plus) for $n_{lq}^+=1$ ($n_{lq}^+=0$), i.e. the associated operator $\hat\sigma_y$ originates from a creation (annihilation) operator.  

Each hermitian operator $\hat V_{lm}$ can always be transformed such that
\begin{equation}
	\hat V_{lm}=\hat R_{lm}^\dagger\hat D_l\hat R_{lm},
	\label{b.20}
\end{equation}
where $\hat R_{lm}$ is a unitary operator and $\hat D_l$ is a real diagonal operator.  Because $\hat V_{lm}$ is a tensor product on $\mathtt H^{\otimes Q}$, we can diagonalize the operator on each space $\mathtt H$ separately.  Using the rotation operators
\begin{equation}
	\begin{array}{cc}
		\hat R_x=\dfrac{\hat\sigma_0-i\hat\sigma_x}{\sqrt{2}},\quad & \quad\hat R_y=\dfrac{\hat\sigma_0+i\hat\sigma_y}{\sqrt{2}},
	\end{array}
	\label{b.21}
\end{equation}
which describe one-qubit rotations around the $x$ and $y$ axes by $\pi/2$ and $-\pi/2$, respectively, we find
\begin{equation}
	\begin{array}{cc}
		\hat\sigma_x=\hat R_y^\dagger\hat\sigma_z\hat R_y,\quad & \quad\hat\sigma_y=\hat R_x^\dagger\hat\sigma_z\hat R_x.
	\end{array}
	\label{b.22}
\end{equation}
Thus, the solution to Eq.\,(\ref{b.20}) is
\begin{subequations}
\begin{align}
	\hat R_{lm}&=\hat\pi_l\bigotimes_{q\in\dot{\mathcal Q}_l}^{\vphantom{}}\hat R_y^{\delta_{m_qx}}\hat R_x^{\delta_{m_qy}}\bigotimes_{q\in\vec{\mathcal Q}_l}^{\vphantom{}}\hat\sigma_0,\label{q.34a}\\
	\hat D_l&=\hat\pi_l\bigotimes_{q\in\dot{\mathcal Q}_l}^{\vphantom{}}\hat\sigma_z\bigotimes_{q\in\vec{\mathcal Q}_l}\hat\sigma_0,\label{q.34b}
\end{align}
	\label{b.23}%
\end{subequations}
for all $m\in\mathcal M_l$.

Inserting Eq.\,(\ref{b.20}) into Eq.\,(\ref{b.17}) and multiplying by Eq.\.(\ref{b.16b}), we eventually find the map
\begin{equation}
	\hat C_l\hat P_{\vec{n}}=\frac{1}{2^{\dot{Q}_l}}\sum_{m\in\mathcal M_l}V_{lm}\hat R_{lm}^\dagger\Big[\sum_{n\in\mathcal N}Z_{n}\hat P_n\Big]\hat R_{lm},
	\label{b.24}
\end{equation}
on $\mathtt H^{\otimes Q}$, where the eigenvalue
\begin{equation}
	Z_{ln} =\prod_{q\in\dot{\mathcal Q}_l}(-1)^{n_q},
	\label{b.25}
\end{equation}
follows from the eigenvalue $(-1)^{n_q}$ of $\hat\sigma_z$ for the state $\ket{n_q}$ on $\mathtt H$.  Inserting Eq.\,(\ref{b.24}) into Eq.\,(\ref{b.14}) finally yields the map
\begin{align}
	e^{-i\hat\lambda^\dagger(\boldsymbol{\theta})}\hat H_le^{i\hat\lambda(\boldsymbol{\theta})} =& \frac{\pi_lh_l}{2^{\dot{Q}_l}}\sum_{n\in\mathcal N}e^{-i\lambda_{\dot{n}_l^+\vec n}^*(\boldsymbol{\theta})}e^{i\lambda_{\dot{n}_l^-\vec n}(\boldsymbol{\theta})}N_{ln}Z_{ln}\nonumber\\
	&\times\sum_{m\in\mathcal M_l}V_{lm}\hat R_{lm}^\dagger\ket{n}\!\bra{n}\hat R_{lm},
	\label{b.26}
\end{align}
on $\mathtt H^{\otimes Q}$.

As any quantum state $\ket{\Psi_0}\in\mathcal F$ maps to $\ket{\Psi_0}\in\mathtt H^{\otimes Q}$, we can now express the expectation value in Eq.\,(\ref{e.5a}), using Eq.\,(\ref{b.2}) and (\ref{b.26}), as Eq.\,(\ref{e.14a}) with the coefficients in Eq.\,(\ref{e.15}) given by Eq.\,(\ref{b.13}), (\ref{b.18}), and (\ref{b.25}) for $N_{ln}$, $V_{lm}$, and $Z_n$, respectively.  As the operator in Eq.\,(\ref{b.8}) is already diagonal, one also straightforwardly finds Eq.\,(\ref{e.14b}) from this equation.

\end{document}